\documentclass[10pt,letterpaper]{article}
\usepackage[utf8]{inputenc}
\usepackage{graphicx}
\usepackage{subfig}
\usepackage{amssymb,amsmath}
\usepackage{authblk}
\usepackage{textcomp}

\def\opex{ Opt.\ Express }

\def\apb{ Appl.\  Phys.\ B }
\def\apl{ Appl.\ Phys.\ Lett.\ }

\def\jmo{ J.\ Mod.\ Opt.\ }

\def\josab{ J.\ Opt.\ Soc.\ Am.\ B }

\def\ol{ Opt.\ Lett.\ }

\def\pra{ Phys.\ Rev.\ A }

\def\prd{ Phys.\ Rev.\ D }

\def\prl{ Phys.\ Rev.\ Lett.\ }

\DeclareMathAlphabet{\mathcal}{OMS}{cmsy}{m}{n}

\graphicspath{{graphics/}}

\newcommand\scalemath[2]{\scalebox{#1}{\mbox{\ensuremath{\displaystyle #2}}}}

\title{Squeezed light in an optical parametric oscillator network with coherent feedback quantum control}   

\author[1]{Orion Crisafulli}
\author[1]{Nikolas Tezak\footnote{ntezak@stanford.edu}}
\author[1,2]{Daniel B. S. Soh}
\author[1]{Michael A. Armen}
\author[1]{Hideo Mabuchi}

\affil[1]{Ginzton Laboratory, Stanford University, 348 Via Pueblo Mall, Stanford, CA 94305} 
\affil[2]{Sandia National Laboratories, 7011 East Avenue, Livermore, CA 94566}

\begin{document}

\maketitle

\begin{abstract}
We present squeezing and anti-squeezing spectra of the output from a degenerate optical parametric oscillator (OPO) network arranged in different coherent quantum feedback configurations. One OPO serves as a quantum plant, the other as a quantum controller.  The addition of coherent feedback enables shaping of the output squeezing spectrum of the plant, and is found to be capable of pushing the frequency of maximum squeezing away from the optical driving frequency and broadening the spectrum over a wider frequency band. The experimental results are in excellent agreement with the developed theory, and illustrate the use of coherent quantum feedback to engineer the quantum-optical properties of the plant OPO output.
\end{abstract}



\section{Introduction}

Coherent feedback quantum control schemes enable the design of autonomous controllers for quantum systems \cite{wiseman1994all, lloyd2000coherent, yanagisawa2003transfer, gough2009series}. Unlike measurement based control schemes, coherent feedback quantum control does not rely on measurement based feedback, which is always accompanied by undesirable side-effects, including the addition of excess noise from amplifying quantum signals up to macroscopic levels and increased controller latency \cite{belavkin1992quantum, bouten2007introduction}. Instead, the feedback signals remain coherent and quantum, which not only reduces excess noise, but has the potential to lower required power for operation, as well as reduce feedback latency and increase system autonomy \cite{mabuchi2011lowpower,kerckoff2010QEC}.

The foundations that underpin the use of coherent feedback to control quantum systems were laid over the past three decades in studies of open quantum systems and networks. The seminal work of Hudson and Parthasarathy \cite{hudson1984quantum} provided a rigorous mathematical framework for describing open quantum systems through quantum stochastic processes \cite{gardiner1985input}. Later work by Gardiner and Carmichael addressed the extension to cascaded quantum systems in \cite{gardiner1985cascaded, carmichael1993cascaded} and more recently Gough and James generalized this to arbitrary feedforward and feedback networks in \cite{gough2009series}. The latter presented a unified theoretical approach to describe the different possible types of coherent connections between systems (including series, concatenation and feedback forms). This approach has significant merit for quantum information science applications as it provides systematic methods to tailor the design of a coherent feedback network to realize a system with desired quantum dynamics. To date only a few experimental demonstrations of such coherent feedback control networks have been reported. Mabuchi experimentally demonstrated disturbance rejection using a dynamic compensator with passive systems (a pair of empty cavities driven by laser fields optically connected in a phase coherent feedback configuration) \cite{mabuchi2008coherent}, and more recently Iida \emph{et al.} experimentally demonstrated optical squeezing enhancement produced from an OPO with self-feedback from its output \cite{iida2012experimental}.  Neither of these demonstrations included quantum-limited \emph{gain} in the arm of the feedback loop, and one of the contributions of the present work is to examine the effects of gain on the feedback system dynamics.
In addition, coherent feedback has been explored in a quantum system with feedback fields in the microwave (as opposed to optical) regime \cite{kerckhoff2012multivibrator}, illustrating that the developed theoretical framework is applicable to quantum systems with significantly different energy and timescales than those of optical systems.

Squeezed states of light find applications in quantum metrology, quantum sensing, and quantum key distribution \cite{ralph2003qkd, vahlbruch2007quantum}. They were theoretically predicted by Collett and Gardiner \cite{collett1984squeezing}, and were observed soon thereafter by Slusher \emph{et al.}  \cite{slusher1985observation}. Wu \emph{et al.} then demonstrated quadrature squeezed light generation from the output of a degenerate OPO \cite{wu1987squeezed}. Quadrature squeezed states play a particularly important role in gravitational wave detection, where they allow measurement sensitivity to be improved beyond the standard quantum limit \cite{caves1981quantum, thorne1987300, kimble2001conversion}. Since gravitational waves have frequencies in the low acoustic range ($\sim$ Hz to kHz), effort in that field has been focused on generating squeezing at these frequency detunings from the optical beam frequency \cite{gea1987squeezed}. Vahlbruch \emph{et al.} utilized two coherent frequency shifted control fields in orthogonal polarization states to stably squeeze the vacuum using an OPO squeezer \cite{vahlbruch2006coherent}, and they subsequently extended the observable frequency range down to 1 Hz by carefully removing interference from surface scatterers \cite{vahlbruch2007quantum}. Other applications, most notably quantum key distribution, benefit from generating squeezing over large bandwidths \cite{cerf2005qkd}.  It is thus desirable to explore the ways in which the bandwidth and the frequency detuning of squeezed light can be engineered for quantum systems that can produce it, and in this work we focus on a network of two OPOs coupled in a coherent feedback configuration.
While it has been shown that cascading parametric oscillators (feeding the output each into the input of the next) can enhance entanglement and squeezing of the final output light \cite{zan2012opacascade}, the merits of arranging the oscillators in a coherent feedback configuration warrant careful study in order to find ways of enhancing the squeezing produced while minimizing the number of oscillators used to achieve it.

In our system, one OPO behaves as a quantum ``plant'' (system to be controlled), and other as a quantum ``controller'' (system that takes the plant output and feeds a modified version of the output, in this case a squeezed state of light, back into the plant to change the plant system dynamics). As the parameters of the feedback loop and controller OPO are varied (including the optical phase accumulation from propagation in the loop, the controller OPO pump power and cavity detuning), the behavior of the combined system departs in a significant way from the that of the individual plant OPO system. In the next section, we present the theory to describe the full OPO network with coherent feedback. We then describe the experimental setup, and the results, namely the spectral features of the combined system output. These spectra show that we are able to choose the controller system parameters to not only enhance the squeezing produced by the plant OPO, but to shift the frequency of maximum squeezing to larger detunings away from the driving optical frequency, as well as broaden the squeezing spectrum.  Systematic spectral shaping through the use of coherent quantum feedback lies at the core of the novelty of this work, and we observe an excellent agreement between the theoretically calculated and experimentally measured squeezing spectra.  We conclude with a discussion of the results and future work.

\section{Theory}\label{chapter:theory}

The model for the OPO network can most easily be derived using Gough and James's SLH formalism \cite{gough2009series}, which allows the derivation a dynamic network model from the knowledge of its individual components and their interconnections.
We are only interested in operating the OPOs below threshold and can thus treat the pump modes of each OPO as classical amplitudes.
In this case, the equations of motion of the OPO signal mode operators are linear, which enables the characterization of the overall network's effect on a coherent input beam through a transfer function. The system's output quantum field features frequency dependent squeezing and its squeezing spectrum can be computed from the transfer function of the OPO network.

The first OPO has two main input-output ports with coupling rates $\gamma_1$ and $\gamma_2$, and is modeled to have three additional, independent loss channels\footnote{These loss channels have only the vacuum as input.} $\gamma_3, \gamma_4, \gamma_L$, bringing its total number of input-output ports five. The second OPO has one main input-output port with rate $\kappa$ and an additional port to serve as combined loss channel with rate $\kappa_L$.

In an interaction picture frame rotating at the external laser frequency, the corresponding SLH models for these systems are then  $\mathbf{G}_P = (\mathbf{S}_P , \mathbf{L}_P , H_P)$ and $\mathbf{G}_C = (\mathbf{S}_C , \mathbf{L}_C, H_C)$ (P: plant, C: controller) with
\begin{eqnarray}
	\mathbf{S}_P &=& \mathbf{I}_5, \hspace{5cm} \mathbf{S}_C = \mathbf{I}_2, \\
	\mathbf{L}_P &=& \left( \begin{array}{c} \sqrt{\gamma_1} a \\ \sqrt{\gamma_2} a \\ \sqrt{\gamma_3} a \\ \sqrt{\gamma_4} a \\ \sqrt{\gamma_L} a \end{array} \right),  \hspace{3.2cm} \mathbf{L}_C = \left( \begin{array}{c} \sqrt{\kappa} b \\ \sqrt{\kappa_L} b \end{array} \right), \\
	H_P &=& \Delta a^\dagger a + \frac{1}{2i} ( \epsilon^* a^2 - \epsilon a^{\dagger 2}), ~~~~~~~H_C = \delta b^\dagger b + \frac{1}{2i} ( \eta^* b^2 - \eta b^{\dagger 2}).
\end{eqnarray}
Here we have introduced the plant and controller cavity detunings from the external driving frequency $\Delta$ and $\delta$, as well as the pump amplitudes $\epsilon$ and $\eta$.

Thanks to the trivial scattering matrices for both OPOs, we can decompose both models into concatenations of single port SLH models for each port $\mathbf{G}_P = \boxplus_{j=1}^5 \mathbf{G}_{Pj}$ and $\mathbf{G}_C = \mathbf{G}_{C1} \boxplus \mathbf{G}_{C2}$. Here, we have the following SLH parameter lists for each single port SLH models.
\begin{align}
	\mathbf{G}_{P1} &= \left( 1,\sqrt{\gamma_1} a,\Delta a^\dagger a + \frac{1}{2i}(\epsilon^* a^2 - \epsilon a^{\dagger 2}) \right), \quad
	\mathbf{G}_{Pj} = (1,\sqrt{\gamma_j} a, 0), ~~ j = 2,3,4,L, \\*[3mm]
	\mathbf{G}_{C1} &= \left( 1,\sqrt{\kappa} b, \delta b^\dagger b + \frac{1}{2i}( \eta^* b^2 - \eta b^{\dagger 2}) \right), \quad
	\mathbf{G}_{C2} = (1, \sqrt{\kappa_L}b,0).
\end{align}

The connection scheme is as follows: The output from the first mirror of the plant OPO is fed into the first mirror of the controller OPO. The connection includes both the phase shift and the linear loss. The output from the first mirror of the controller OPO is fed into the second mirror of the plant OPO. This connection should also include the linear loss as well as the phase shift. It turns out that we can consolidate the phase shift into one component, rather than using two separate phase shifting components. Let us now consider an ideal, lossless connection. In a decomposed notation, this connection can be written as \cite{gough2009series}
\begin{equation}
	(\mathbf{G}_{P2} \lhd \mathbf{G}_\phi \lhd \mathbf{G}_{C1} \lhd \mathbf{G}_{P1}) \boxplus \mathbf{G}_{P3} \boxplus \mathbf{G}_{P4} \boxplus \mathbf{G}_{PL} \boxplus \mathbf{G}_{C2},
\end{equation}
where the phase shifting system should be represented as $\mathbf{G}_\phi = (e^{i \phi},0,0)$.

However, a realistic system must include the losses in the connections. The loss can be modeled using the beam splitter model
\begin{equation}
	\mathbf{G}_{L_j} = \left( \left( \begin{array}{cc} \alpha_j & \beta_j \\ -\beta_j & \alpha_j \end{array} \right) , \mathbf{0}, 0 \right),
\end{equation}
where we must introduce the second idle channel.
The transmission is defined by $\alpha_j$ parameter. Since we have an explicit component to model phases accumulated along the beam path, we choose $\alpha_j$ positive, and it is then given by $\alpha_j = \sqrt{1 - l_j}$ where $l_j$ is the relative power loss. Then, we must have $|\beta_j| = \sqrt{l_j}$ for unitarity of the $\mathbf{S}$ matrix and we also pick these parameters as real and positive.

We account for propagation losses between the first mirror of the plant OPO to the first mirror of the controller OPO ($l_1$), between the first mirror of the controller OPO and the second mirror of the plant OPO ($l_2$), and finally, between the second mirror of the plant OPO and the homodyne detector ($l_3$).

The entire system then has a total of $n:=8$ pairs of input-output channels, five from the plant, two from the controller, three idle ports due to the losses and minus two, due to the feedback connection. Therefore, the system $\mathbf{S}$ matrix should be an $8 \times 8$ matrix and the system $\mathbf{L}$ should be an $8 \times 1$ vector operator.
Following the rules for computing series and concatenation products in \cite{gough2009series}, we can obtain the whole system SLH model:
\begin{eqnarray}
	\mathbf{S} &=& \left( \scalemath{0.8}{ \begin{array}{cccccccc} e^{i\phi} \alpha_1 \alpha_2 \alpha_3 & -e^{i\phi} \beta_1 \alpha_2 \alpha_3 & - \beta_2 \alpha_3 & - \beta_3 & 0 & 0 & 0 & 0 \\
	e^{i \phi} \beta_3 \alpha_1 \alpha_2 & - e^{i \phi} \beta_1 \beta_3 \alpha_2 & - \beta_2 \beta_3 & \alpha_3 & 0 & 0 & 0 & 0 \\
	e^{i \phi} \beta_2 \alpha_1 & - e^{i \phi} \beta_1 \beta_2 & \alpha_2 & 0 & 0 & 0& 0& 0 \\
	\beta_1 & \alpha_1 & 0 & 0 & 0 & 0 & 0 & 0 \\
	0 & 0 & 0 & 0 & 1 & 0 & 0 & 0 \\
	0 & 0 & 0 & 0 & 0 & 1 & 0 & 0 \\
	0 & 0 & 0 & 0 & 0 & 0 & 1 & 0 \\
	0 & 0 & 0 & 0 & 0 & 0 & 0 & 1	\end{array}} \right), \\*[5mm]
	\mathbf{L} &=& \left( \scalemath{0.95}{ \begin{array}{c} (\sqrt{\gamma_1} \alpha_1 \alpha_2 \alpha_3 e^{i\phi} + \sqrt{\gamma_2} \alpha_3 ) a + \sqrt{\kappa} \alpha_2 \alpha_3 e^{i \phi} b \\
	(\sqrt{\gamma_1} \beta_3 \alpha_1 \alpha_2 e^{i\phi} + \sqrt{\gamma_2} \beta_3)a + \sqrt{\kappa} \beta_3 \alpha_2 e^{i \phi} b \\
	\sqrt{\gamma_1} \beta_2 \alpha_1 e^{i\phi} a + \sqrt{\kappa} \beta_2 e^{i\phi} b	 \\
	\sqrt{\gamma_1} \beta_1 a \\
	\sqrt{\gamma_3} a\\
	\sqrt{\gamma_4} a \\
	\sqrt{\gamma_L} a\\
	\sqrt{\kappa_L} b \end{array}} \right), \label{eq:L} \\*[5mm]
	H&=& \left( \Delta + \sqrt{\gamma_1 \gamma_2}\alpha_1\alpha_2 \sin \phi \right) a^\dagger a + \delta b^\dagger b \nonumber \\
	&& + \frac{1}{2i} \left( \eta^* b^2 - \eta b^{\dagger 2} \right)+ \frac{1}{2i} \left( \epsilon^* a^2 - \epsilon a^{\dagger 2} \right) \nonumber \\
	&& \scalemath{0.85}{+\frac{\sqrt{\kappa}}{2i} \left[ \left( \sqrt{\gamma_2 }\alpha_2 e^{i\phi} - \sqrt{\gamma_1}\alpha_1 \right) ab^\dagger  - \left( \sqrt{\gamma_2}\alpha_2 e^{-i\phi} - \sqrt{\gamma_1}\alpha_1 \right) a^\dagger b \right]}.
\end{eqnarray}
The system Hamiltonian which is still quadratic in the plant and controller mode operators shows that the feedback achieves two things: it introduces a detuning on the plant cavity mode $a$ and it couples the plant and controller modes. The strength of these effects depends on the feedback roundtrip phase, the linear propagation losses, and the coupling coefficients of the plant mirrors 1 and 2.


From the SLH parameters we can compute the QSDEs for the mode dynamics $X\in\{a,a^\dagger, b, b^\dagger\}$:
\begin{equation}
  d X(t) = ( \mathcal{L}(X(t)) - i [ X(t), H]) {\rm dt} + \mathbf{dA}^\dagger(t) \mathbf{S}^\dagger [X(t),\mathbf{L}(t) ] + [\mathbf{L}^\dagger,X(t)] \mathbf{S}\, \mathbf{dA} (t),
\end{equation}
where $\mathbf{dA}(t) = (dA_1(t),dA_2(t),\dots,dA_8(t))^T$ and $\mathbf{dA}^\dagger(t)=(dA_1^\dagger(t),dA_2^\dagger(t),\dots,dA_8^\dagger(t))$ are the input field or more precisely Quantum Ito-type input processes to the system\footnote{which in our case represent either the vacuum or a coherently displaced vacuum} and the Heisenberg-picture Lindblad superoperator $\mathcal{L}$ is given by
\begin{equation}
  \mathcal{L} (X) = \frac{1}{2} \mathbf{L}(t)^\dagger[X,\mathbf{L}(t)] + \frac{1}{2} [\mathbf{L}^\dagger(t),X] \mathbf{L}(t).
\end{equation}
The output field is determined through the boundary condition
\begin{equation}
    \mathbf{dA}'(t) = \mathbf{S}\,\mathbf{dA}(t) + \mathbf{L}(t){\rm dt}.
\end{equation}
Both the QSDEs for the internal modes and the boundary condition for the output field are linear in the internal mode operators and the input fields (and their adjoints), and can thus be represented in the $\mathbf{ABCD}$ matrix formalism \cite{zhang2012qfnreview}.
Here we present only the final result
\begin{eqnarray}
	\mathbf{d\breve{a}}(t) &=& \mathbf{A}\, \breve{\mathbf{a}}(t){\rm dt} + \mathbf{B}\, \mathbf{d\breve{A}}(t), \label{eq:ABCDa}\\
	\mathbf{d\breve{A}}'(t) &=& \mathbf{C}\, \breve{\mathbf{a}}(t){\rm dt} + \mathbf{D}\, \mathbf{d\breve{A}}(t), \label{eq:ABCDb}
\end{eqnarray}
where we use the doubled-up notation of Zhang et al. \cite{zhang2012qfnreview}
\begin{equation}
	\breve{\mathbf{a}} = \left( \begin{array}{c} a \\ b \\ a^\dagger \\ b^\dagger \end{array} \right), ~~ \mathbf{d\breve{A}} = \left( \begin{array}{c} \mathbf{dA} \\ \mathbf{dA}^* \end{array} \right), ~~ \mathbf{d\breve{A}}' = \left( \begin{array}{c} \mathbf{dA}' \\ \mathbf{dA'}^* \end{array} \right).
\end{equation}
With notation borrowed from the above reference $\breve{\Delta} (\mathbf{X}, \mathbf{Y}) \equiv \left( \begin{array}{cc} \mathbf{X} & \mathbf{Y} \\ \mathbf{Y}^* & \mathbf{X}^* \end{array} \right)$, the final system matrices can be expressed as
\begin{equation}
	\mathbf{A} = \breve{\Delta}(\mathbf{A}_-,\mathbf{A}_+), ~~ \mathbf{B} \equiv - \breve{\Delta} ( \mathbf{C}_-^\dagger , \mathbf{0} ), ~~ \mathbf{C} = \breve{\Delta} (\mathbf{C}_- , \mathbf{0} ) , ~~ \mathbf{D} \equiv \breve{\Delta} (\mathbf{S} , \mathbf{0}),
\end{equation}
where
\begin{eqnarray}
	\mathbf{A}_- &=& \left( \begin{array}{cc} - i \Delta - \frac{\gamma_T}{2}- \Gamma e^{i \phi} & - \sqrt{\kappa \gamma_2} \alpha_2 e^{i \phi} \\ - \sqrt{\gamma_1 \kappa}\alpha_1 & - i \delta - \frac{\kappa_T}{2} \end{array} \right), ~ ~
	\mathbf{A}_+ \; = \; \left( \begin{array}{cc} \epsilon & 0 \\ 0 & \eta \end{array} \right), \\
    \mathbf{C}_- &=& \left( \scalemath{0.95}{ \begin{array}{cc} \sqrt{\gamma_1} \alpha_1 \alpha_2 \alpha_3 e^{i\phi} + \sqrt{\gamma_2} \alpha_3 &  \sqrt{\kappa} \alpha_2 \alpha_3 e^{i \phi} \\
        \sqrt{\gamma_1} \beta_3 \alpha_1 \alpha_2 e^{i\phi} + \sqrt{\gamma_2} \beta_3 & \sqrt{\kappa} \beta_3 \alpha_2 e^{i \phi} \\
        \sqrt{\gamma_1} \beta_2 \alpha_1 e^{i\phi} & \sqrt{\kappa} \beta_2 e^{i\phi}    \\
        \sqrt{\gamma_1} \beta_1 & 0 \\
        \sqrt{\gamma_3} & 0\\
        \sqrt{\gamma_4} & 0\\
        \sqrt{\gamma_L} & 0\\
        0 &\sqrt{\kappa_L}  \end{array}} \right) \\
	\Gamma &=& \sqrt{\gamma_1 \gamma_2} \alpha_1\alpha_2, \\
	\gamma_T &=& \gamma_1 + \gamma_2 + \gamma_3 + \gamma_4 + \gamma_L, \\
	\kappa_T &=& \kappa + \kappa_L.
\end{eqnarray}

By Fourier transforming equations \eqref{eq:ABCDa} and \eqref{eq:ABCDb}, we can find the system transfer function $\Xi$ that directly links the Fourier transformed input and output fields $\mathbf{\tilde{\breve{A}}}^{(')}(\omega):=\frac{1}{2\pi}\int_{-\infty}^\infty e^{i \omega t} \mathbf{d\breve{A}}^{(')}(t)$ as
\begin{equation}
	\mathbf{\tilde{\breve{A}}}' (\omega) = \Xi (\omega) \mathbf{\tilde{\breve{A}}} (\omega),
\end{equation}
where
\begin{equation}
	\Xi (\omega) \equiv \mathbf{D} + \mathbf{C} ( - i \omega \mathbf{I}_{4} - \mathbf{A} )^{-1} \mathbf{B}.
\end{equation}
Due to the inherent redundancy of the doubled up notation, the transfer function can be decomposed as
\begin{equation}
	\Xi (\omega) = \left( \begin{array}{cc} \mathcal{S}^- (\omega) & \mathcal{S}^+ (\omega) \\ {\mathcal{S}^+}^* (-\omega) & {\mathcal{S}^-}^* (-\omega) \end{array} \right).
\end{equation}
From these matrices one can calculate the quadrature dynamics and eventually calculate the power spectral density of a given quadrature, which is known as squeezing spectrum \cite{gough2009enhancement}
\begin{equation}
	\mathcal{P}_j^\theta (\omega) = 1 + \mathcal{N}_j (\omega) + \mathcal{N}_j (-\omega) + e^{2 i \theta} \mathcal{M}_j (\omega) + e^{- 2 i \theta} \mathcal{M}_j (-\omega),
\end{equation}
for $j=1,2,...$ representing each output port and $\theta$ as the quadrature angle. In our case, the feedback system output is $j = 1$. Here, the parameters $\mathcal{N}_j (\omega)$ and $\mathcal{M}_j (\omega)$ are defined through
\begin{equation}
	\mathcal{N}_j (\omega) \equiv \sum_k | \mathcal{S}_{jk}^+ (\omega) |^2, ~~ \mathcal{M}_j (\omega) \equiv \sum_k S_{jk}^- (\omega) \mathcal{S}_{jk}^+ (-\omega),
\end{equation}
where $\mathcal{S}_{jk}^-$ and $\mathcal{S}_{jk}^+$ are the $(j,k)$ entries of the matrices $\mathcal{S}^- (\omega)$ and $\mathcal{S}^+(\omega)$, respectively.

So far, the expressions for the squeezing spectrum apply only to vacuum inputs. If we displace the input by a vector of coherent amplitudes $\mathbf{dA}(t)\rightarrow \mathbf{dA}(t) + \mathbf{w}\, {\rm dt}$ the squeezing spectrum is only altered at dc ($\omega = 0$) in the form of a singular $\propto\delta(\omega)$ contribution:
\begin{equation}
	\mathcal{P}_j^\theta (\omega) \rightarrow \mathcal{P}_j^\theta (\omega) +
4 \left\{{\rm Re}\left[e^{i\theta}v_j(\mathbf{w})\right]\right\}^2  \delta (\omega),
\end{equation}
where we have introduced the $j$-th steady state output amplitude $v_j(\mathbf{w})$, given by
\begin{equation}
	v_j (\mathbf{w} ) := \left(\Xi(0)\mathbf{\breve{w}}\right)_j = \left(\mathcal{S}^-(0)\mathbf{w} + \mathcal{S}^+(0)\mathbf{w}^*\right)_j.
\end{equation}

The exact analytical form of the squeezing spectrum we obtained is too complicated to be shown here and is visually uninformative, but we were able to numerically evaluate the squeezing spectrum for given system parameters, which were readily compared with the experimental results. The following sections show the comparison between the theoretically calculated squeezing spectra and the experimental results.

For completeness, since this is relevant to our locking schemes, we also include how to compute the steady state output power from a given port. For the $j$-th port, this is given by the steady state expectation of the output gauge process matrix element $P_j^{\rm ss} {\rm dt} = \langle d\Lambda_{jj}'\rangle_{\rm ss}$ and can be shown to be
\begin{align}
P_j^{\rm ss} = \left|v_j(\mathbf{w})\right|^2 + \left(\mathbf{C}\Delta \mathbf{N}_{\rm ss}\mathbf{C}^\dagger\right)_{n+j, n+j},
\end{align}
where the first term is simply the squared output amplitude and the second arises purely due to the squeezed uncertainty ellipses of the intra-cavity modes.
The intra-cavity modes' covariance matrix $\Delta \mathbf{N}_{\rm ss} := \left \langle \left(\mathbf{\breve{a}} -  \left\langle \mathbf{\breve{a}} \right\rangle_{\rm ss} \right)\left(\mathbf{\breve{a}} -  \left\langle \mathbf{\breve{a}} \right\rangle_{\rm ss} \right)^\dagger \right \rangle_{\rm ss} = \langle\mathbf{\breve{a}}\mathbf{\breve{a}}^\dagger\rangle_{\rm ss} - \langle\mathbf{\breve{a}}\rangle_{\rm ss} \langle\mathbf{\breve{a}}^\dagger\rangle_{\rm ss}$ is obtained by solving the following Liapunov equation:
\begin{align}
    \mathbf{0} & = \mathbf{A} \Delta\mathbf{N}_{\rm ss} +\Delta\mathbf{N}_{\rm ss}\mathbf{A}^\dagger
 +  \mathbf{Q},
\quad\text{with }\mathbf{Q} := \mathbf{B}\frac{\langle\mathbf{d\breve{A}}\mathbf{d\breve{A}}^\dagger\rangle}{{\rm dt}}\mathbf{B}^\dagger = \begin{pmatrix}
 \mathbf{C}_-^\dagger\mathbf{C}_- & \mathbf{0} \\
 \mathbf{0} & \mathbf{0}
 \end{pmatrix}
\end{align}	
	
\section{Experimental setup}

\begin{figure}
	\centering
	\includegraphics[width = 0.88\textwidth]{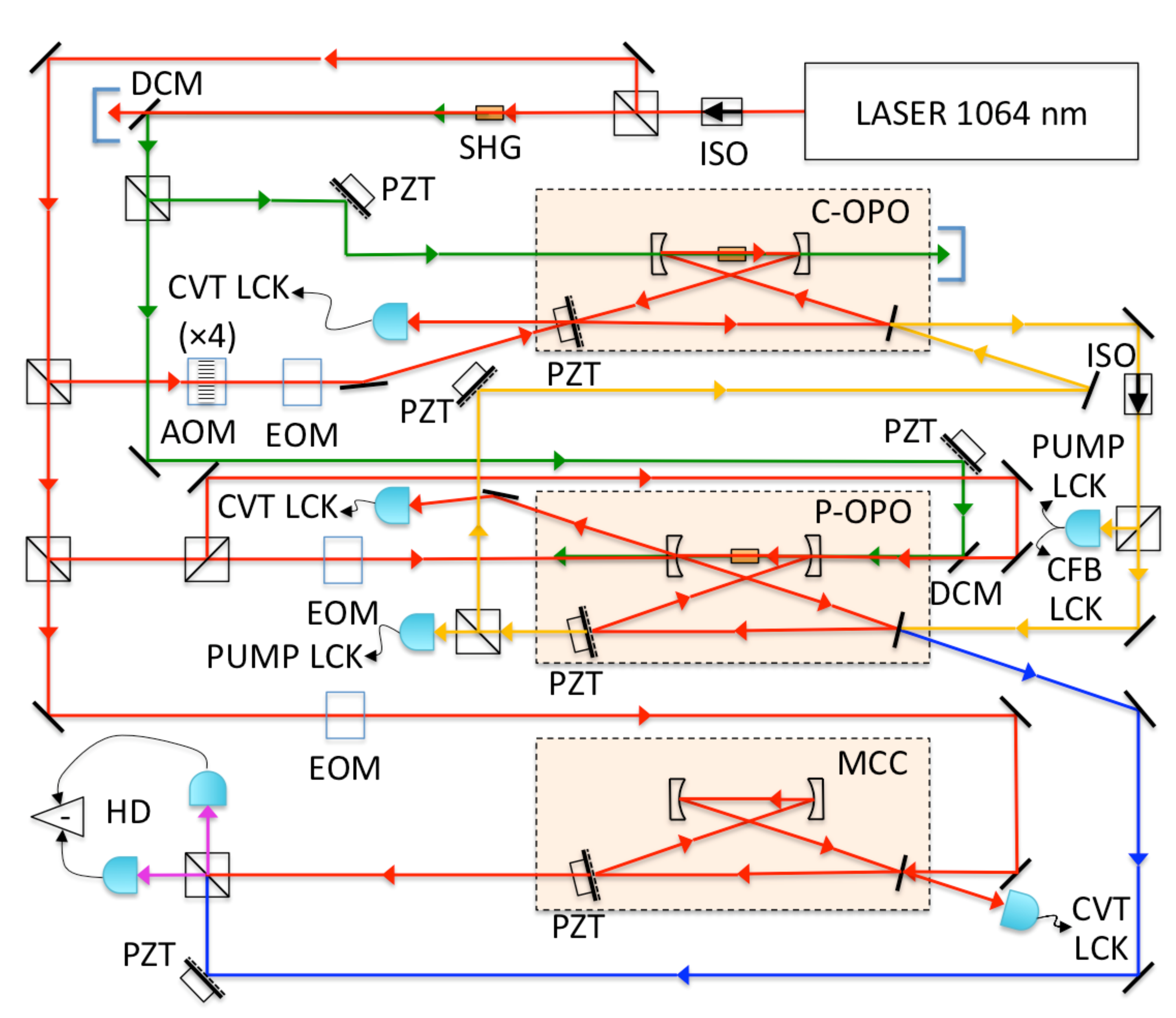}
	\caption{Schematic of the OPO network system. Red: 1064 nm, green: 532 nm, orange: coherent feedback, blue: system output. (ISO: isolator, SHG: second-harmonic generation, DCM: dichroic mirror, PZT: piezoelectric transducer, C-OPO: controller OPO, P-OPO: plant OPO, MCC: mode-cleaning cavity, CVT LCK: cavity locking servo, AOM: acousto-optic modulator, EOM: electro-optic modulator, PUMP LCK: pump phase locking servo, CFB LCK: coherent feedback phase locking servo, HD: Homodyne detection) }\label{fig:system_scheme}
\end{figure}

The system consists of two OPOs in a feedback loop. Figure \ref{fig:system_scheme} shows the entire system diagram. We have three optical cavities. The output of the plant OPO is connected to the controller OPO input and the controller output is fed back into another input port of the plant OPO. This forms an all-optical coherent feedback quantum control system. The master laser source for the entire system is a single frequency CW 1064 nm laser (Mephisto MOPA, InnoLight), operating at 1064.4 nm. We generate a 532 nm pump beam through single pass second harmonic generation (SHG) using a MgO:PPLN crystal (Covesion MSHG1064-1.0-10 with a 1 mm $\times$ 1 mm aperture and 1 cm length) with a poling period of 6.96 \textmu m, set at a temperature of 59.4 $^\circ$C. The maximum 532 nm power obtained is 1.7 W when the pump power is 9.6 W. Part of the master laser is split off for use as a local oscillator for the balanced homodyne detector. Since the local oscillator beam quality is not ideal, we employ a mode cleaning cavity. This cavity is locked using the Pound-Drever-Hall (PDH) method \cite{drever1983laser}. For this, we generate the sideband of the locking laser beam at 97 MHz. This mode cleaning cavity also serves to stabilize the local oscillator power.

We configure our plant OPO to achieve degenerate three-wave mixing, which is pumped by the 532 nm beam. The plant cavity is in a bow-tie configuration, with two input-output mirrors. The mirrors have 8\% and 4\% reflectivity, respectively. These values were chosen to provide reasonable coupling strength to the plant cavity in the feedback loop. The plant cavity also has two concave mirrors with 99.66\% and 99.90\% reflectivity at 1064 nm, but with reasonably high transmission ($>$93\%) at 532 nm. The radius of curvature for both mirrors is 10 cm. The cavity round-trip length is approximately 67 cm. The major intracavity loss occurs at the surface of the MgO:PPLN, which has a non-ideal anti-reflection coating with approximately 1\% loss. Locking the plant OPO cavity is also achieved using the PDH method. For this we control the upper flat mirror position using a piezoelectric transducer (PZT). The cavity locking beam is coupled through one of the concave mirrors in the backward direction. The PDH sideband frequency is 90.7 MHz, which is much larger than the cavity half-linewidth of 5.2 MHz. Although we employ the backward beam cavity locking scheme, locking is still challenging due to significant backscattering of the relatively strong forward signal on the MgO:PPLN surfaces.
The plant OPO is a type I degenerate OPO and uses the same type of MgO:PPLN crystal as the one used for SHG. The pump beam is coupled through the lower concave mirror in the forward direction, co-propagating with the signal. The desired pump beam parameters for optimizing parametric amplification were determined by matching the pump beam parameters to those of the second harmonic beam generated in the plant OPO from the forward signal power. The poling period of this MgO:PPLN crystal is also 6.96 \textmu m, and the crystal is mounted in an oven with a set temperature of 35.6 $^\circ$C. We measure the OPO oscillation threshold pump power for the plant OPO to be 330 mW. The pump phase relative to the signal phase is locked through a dithering scheme analogous to the PDH method with the modulation sideband at 111.1 kHz. We monitor the output of the plant OPO using a 5\% beam splitter to generate the error signal used in the dithering scheme; although this adds loss in the feedback loop, it greatly improves the pump phase locking performance.

The controller OPO is also a bow-tie cavity. It has a single main input-output mirror with 86\% reflectivity. All other mirrors are high reflectors. It also uses the same type of MgO:PPLN crystal as the plant OPO, has a poling period of 6.93 \textmu m, and the crystal sits in an oven with a set temperature of 58.1$^\circ$C.
Although the controller OPO is designed as an ideal single input-output port OPO to produce as much squeezing as possible, we found that additional intracavity loss caused by the MgO:PPLN surface and absorption loss in the crystal degrades the on-resonance cavity reflectivity at the input-output port from the near ideal value of 100\% to 70\%. This corresponds to a 1.3\% intracavity loss. Locking the controller OPO cavity by injecting a backward beam is challenging for several reasons: the remaining cavity mirrors are high reflectors and let little backward beam power through, and the small reflection of the forward signal from the PPLN crystal surface at the same optical frequency can significantly interfere with the injected backward beam. To avoid this problem we shift the optical frequency of the cavity locking beam by precisely one cavity free-spectral range (435 MHz), and we achieve this frequency shift by passing the cavity locking beam through an AOM four times before it is injected into the cavity. We lock the cavity using the PDH method with a modulation sideband at 48 MHz. Locking the pump phase relative to the signal phase is accomplished using through the same low frequency dithering scheme as that used in the plant OPO, and the error signal is generated from a detector that monitors the output of the controller OPO through a 10\% beam splitter located near the input of the plant OPO. Just as with the plant OPO, this greatly enhances our ability to lock the controller pump phase, at the expense of additional loss in the feedback loop. We measure the controller OPO threshold to be 993 mW.

The total roundtrip phase of the feedback loop is locked in either constructive interference (maximizing the circulating power in the feedback loop) or destructive interference (minimizing the circulating power). For this we seed a small amount of forward signal power to the plant OPO and then monitor the circulating power through the same 10\% beam splitter used for locking the controller OPO pump phase. We have another detector at the far end of the feedback loop that then enables us to monitor the beam alignment drift at a separate position.

One of the relay feedback loop mirrors serves to adjust the feedback loop phase. The minimum power monitored corresponds to $P_0 / |1 + |r||^2$, where $r$ is the complex loop gain, while the maximum power corresponds to $P_0 / |1 - |r||^2$. We easily measure $P_0$ by cutting the loop. From these three measurements, we obtain $|r| = 0.6$ for cold OPOs (no pump), which implies that the total loop power transmission was $0.6^2 = 0.36$ (i.e., 64\% loop power loss). Through pointwise measurements, we break down the contributions as follow: 10\% and 5\% losses due to the monitoring taps, the cavity reflectivity of 70\% of the locked controller OPO, 12\% of loss from an optical isolator (which is used to ensure uni-directional feedback), the cavity transmissivity 82\% of the locked plant OPO, and finally, additional losses ($\sim$ 7\%) from mode coupling mismatch and various mirror surfaces, including 14 dielectric coated mirrors for positioning and aligning beams, 4 lenses to shape the beams, and one silver mirror on a piezo-electric material. The theory predicts only slight squeezing spectrum enhancements for significantly reduced feedback loss, and thus minimizing feedback loop loss is not crucial for the results demonstrated here.

To monitor squeezing spectra we set up a balanced homodyne detection scheme \cite{yuen1980optical,yuen1983noise,breitenbach1997measurement}. The output from the mode cleaning cavity serves as a local oscillator, which is mixed with the output signal from the plant OPO on a beam splitter. The detectors we use are two ETX-500T photodiodes (JDSU), arranged in subtracting configuration. The total losses in the homodyne measurement setup can be decomposed into three main components: the propagation loss from the output of the plant OPO to the surface of the detectors, the homodyne efficiency measured from the homodyne visibility fringe, and the quantum efficiency of the detector \cite{takeno2007observation}. We have an overall loss of 30\% in the homodyne measurement including all three components. Locking the local oscillator phase relative to the signal phase is accomplished through the same dithering scheme mentioned above, and dithering occurs at 373.5 kHz, which is well separated from the other phase lock loop dithering frequencies. This helps to lock the homodyne quadrature angle effectively while all seven phase locking loops are running simultaneously, at the expense of adding excess technical noise in the homodyne spectra at frequencies in the vicinity of this dithering frequency.

Locking the entire system requires a sequential procedure, due to interaction among the various phase lock loops. In sequence, we first lock the mode cleaning cavity, plant OPO cavity, controller OPO cavity, the feedback phase, and the homodyne local oscillator phase. Then, we lock the pump phases while monitoring the homodyne output, first the plant OPO, and then, finally, the controller OPO. We lock the balanced homodyne output at either maximum or at minimum, representing the amplitude quadratures ($\theta = 0, \pi$). To change between measuring squeezing and anti-squeezing, we switch the OPO pump phases relative to the seeding input.

The squeezing spectrum is taken directly from the output of the balanced homodyne detector. The output is connected to a Minicircuits bias tee splitter which separates the signal into DC and AC components, and the measured squeezing spectra are contained in the AC component; the DC component is used to monitor the DC bias drift of the balanced homodyne detector, which is caused by the slow beam drift and the resulting imbalance in measured power of two detectors.

\section{Experimental results}

\begin{figure}
	\centering
	\includegraphics[width=0.6\textwidth]{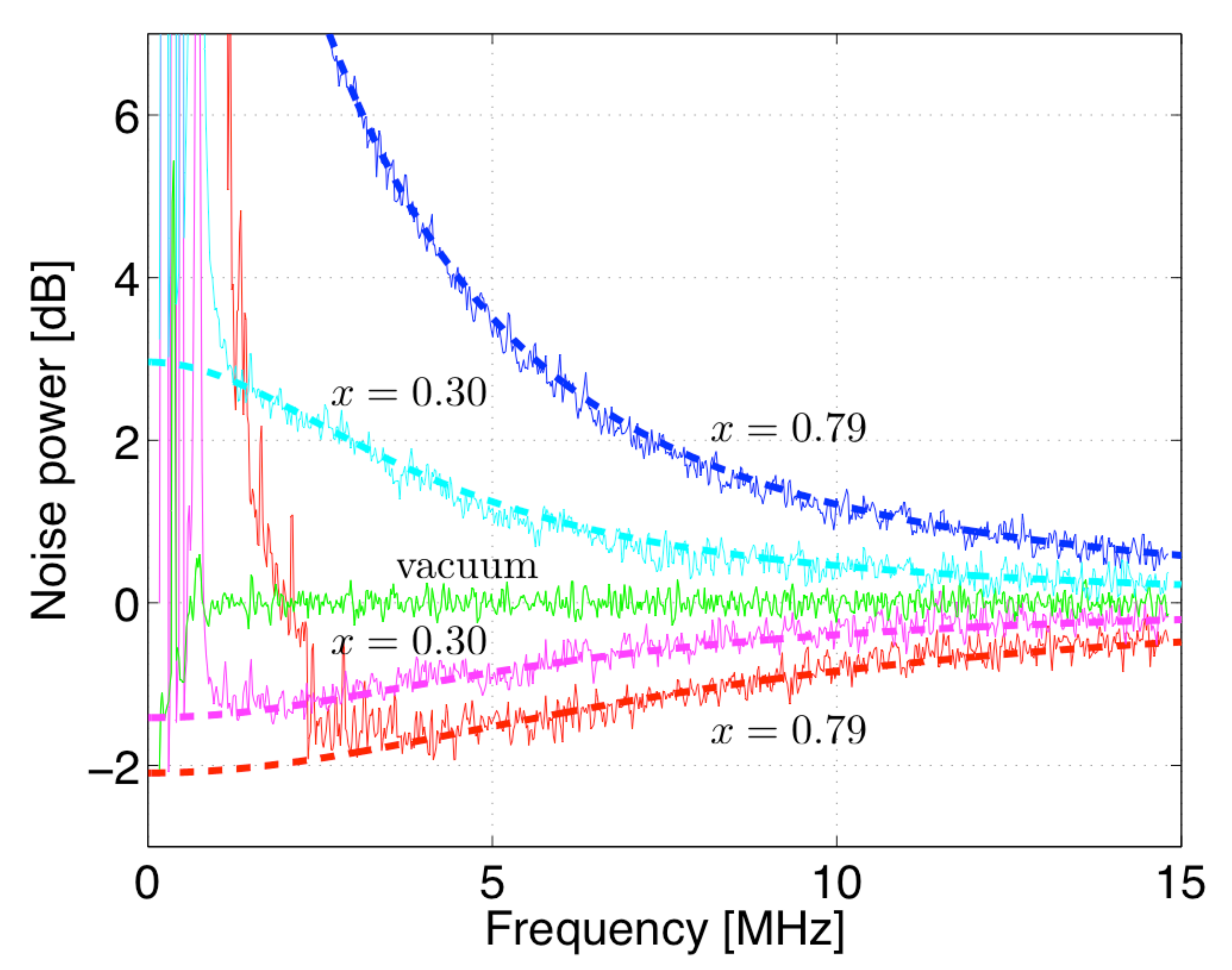}
	\caption{Squeezing spectra of the plant OPO without feedback at two different pump powers. Blue (cyan): anti-squeezing, red (magenta): squeezing for $x = 0.79$ ($x = 0.3)$. Here, $x = \sqrt{P^P_p / P^P_\text{th}}$. Green: vacuum spectrum. Theoretical curves are shown in dashed lines.}\label{fig:measured-spectra}
\end{figure}

We first measure the squeezing spectra of the plant OPO without feedback using a RF spectrum analyzer (Agilent 4396A) with 30 kHz RBW and 300 Hz VBW, averaged 9 times. The pump parameter is defined as $x \equiv 2 \epsilon / \gamma_T = \sqrt{P^P_p / P^P_\text{th}}$, where $P^P_p$ is the plant pump power and $P^P_\text{th}$ is the plant threshold pump power. For all measurements, the contributions of the photodetector dark current and the small spectral variation of the transfer function of the RF spectrum analyzer are accounted for. The local oscillator power is 0.25 mW and we maintain the forward beam signal power at roughly 2 \textmu W, which avoids saturation of the homodyne detector while providing sufficiently large optical power to lock various phase locking servos and enables large dynamic range in the spectrum. To measure squeezing (and anti-squeezing), we locked the pump phase of the plant OPO for deamplification (and amplification), respectively. Figure \ref{fig:measured-spectra} shows the measured spectra. We calculate the theoretical curves using the parameters shown in the table \ref{tab:param-emptycavity}. For the open loop calculation, the feedback loss is 100\%. We observe that the spectrum data from 0 Hz to 2.5 MHz is obscured by noise from the dithering RF locking signals, whose harmonics and spectral tails die out above 2.5 MHz.

\begin{table}[!htb]
	\centering
	\begin{tabular}{c  c  c}
        \hline
        \hline
        \\*[-2mm]
		Symbol & Value & Description \\*[3mm]
		\hline \\
		$\gamma_1$ & $4 \times 4.5 \text{MHz}/2\pi= 18 \text{MHz}/2 \pi$ & Plant mirror 1 \\
		$\gamma_2$ & $8 \times 4.5 \text{MHz}/2\pi=36 \text{MHz}/2 \pi$ & Plant mirror 2 (main output of plant OPO) \\
		$\gamma_3$ & $.34 \times 4.5 \text{MHz}/2\pi \approx 2 \text{MHz}/2 \pi$ & Plant mirror 3 (concave mirror) \\
		$\gamma_4$ & $.1 \times 4.5 \text{MHz}/2\pi \approx .45 \text{MHz}/2 \pi$ & Plant mirror 4 (concave mirror) \\
		$\gamma_L$ & $2 \times 4.5 MHz/2\pi = 9 \text{MHz}/2 \pi$ & Combined additional Plant losses \\
		$\Delta$   & $0 \text{MHz}/2\pi$ & Plant cavity detuning \\
		\hline \\*[-3mm]
		$\kappa$ & $14 \times 4.34 \text{MHz}/2\pi \approx 61 \text{MHz}/2\pi$ & Controller input-output mirror \\
		$\kappa_L$ & $1.3 \times 4.34 \text{MHz}/2\pi \approx 5.7 \text{MHz}/2\pi$ & Extra controller intracavity loss \\
		$\eta$ & $0 \text{MHz} /2\pi$ & Controller pump power gain \\
		$\delta$ & $0 \text{MHz}/2 \pi$ & Controller cavity detuning \\
		\hline \\*[-3mm]
		$\phi$ & $\pi$ & Feedback optical phase (destructive) \\
		$l_1$ & 3.5\% & Power loss from plant to controller \\
		$l_2$ & 27\% & Power loss from controller to plant \\
		$l_3$ & 30\% & Homodyne power loss \\*[3mm]
		\hline
        \hline
	\end{tabular}
	\caption{Parameters used to fit measured squeezing spectra of the empty-cavity destructive interference coherent feedback configuration.}\label{tab:param-emptycavity}
\end{table}	

\begin{figure}[!tb]
	\centering
	\subfloat[$x = 0.17$ ]{\includegraphics[width=0.5\textwidth]{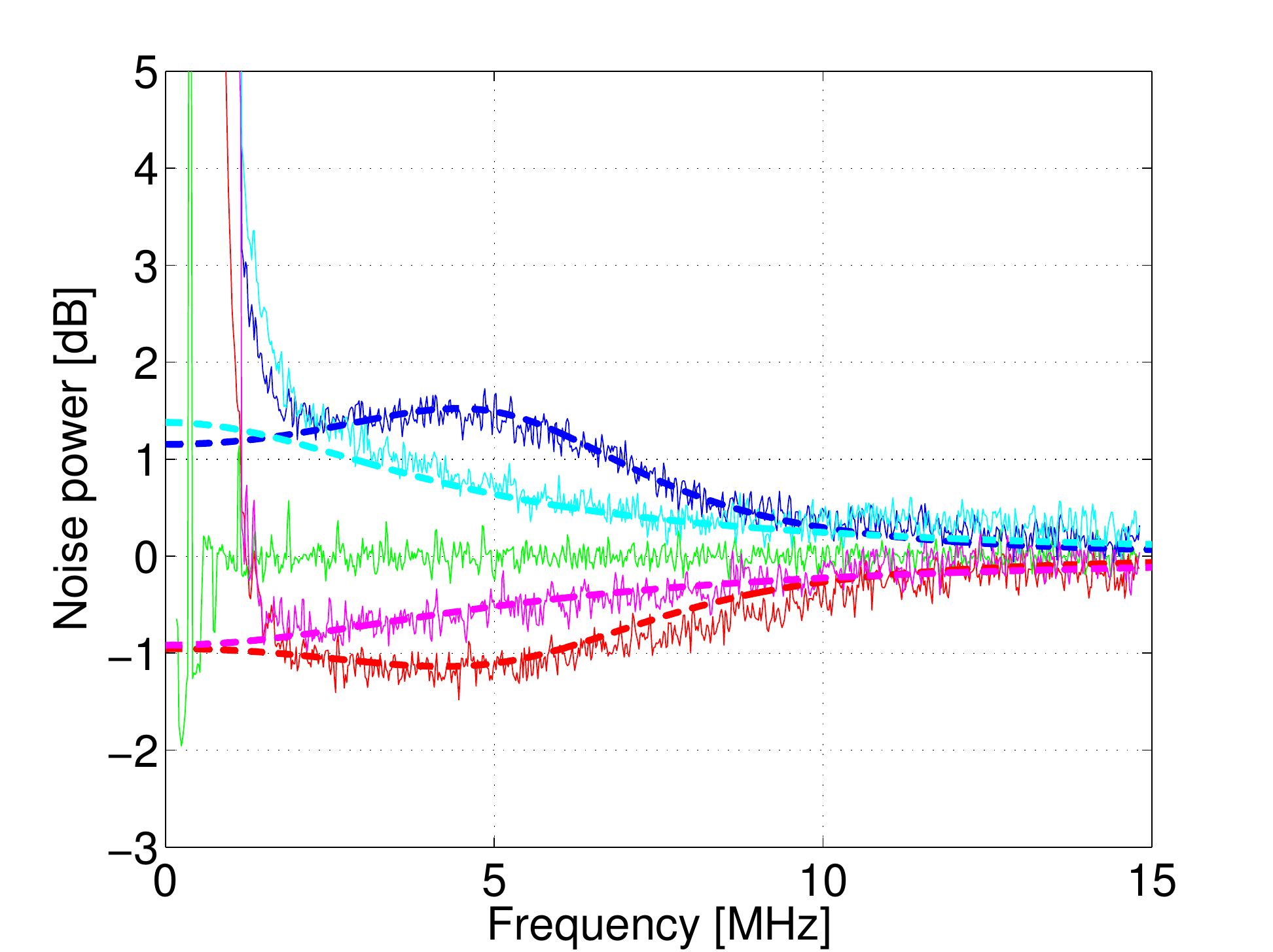}}
	\subfloat[$x = 0.33$ ]{\includegraphics[width=0.5\textwidth]{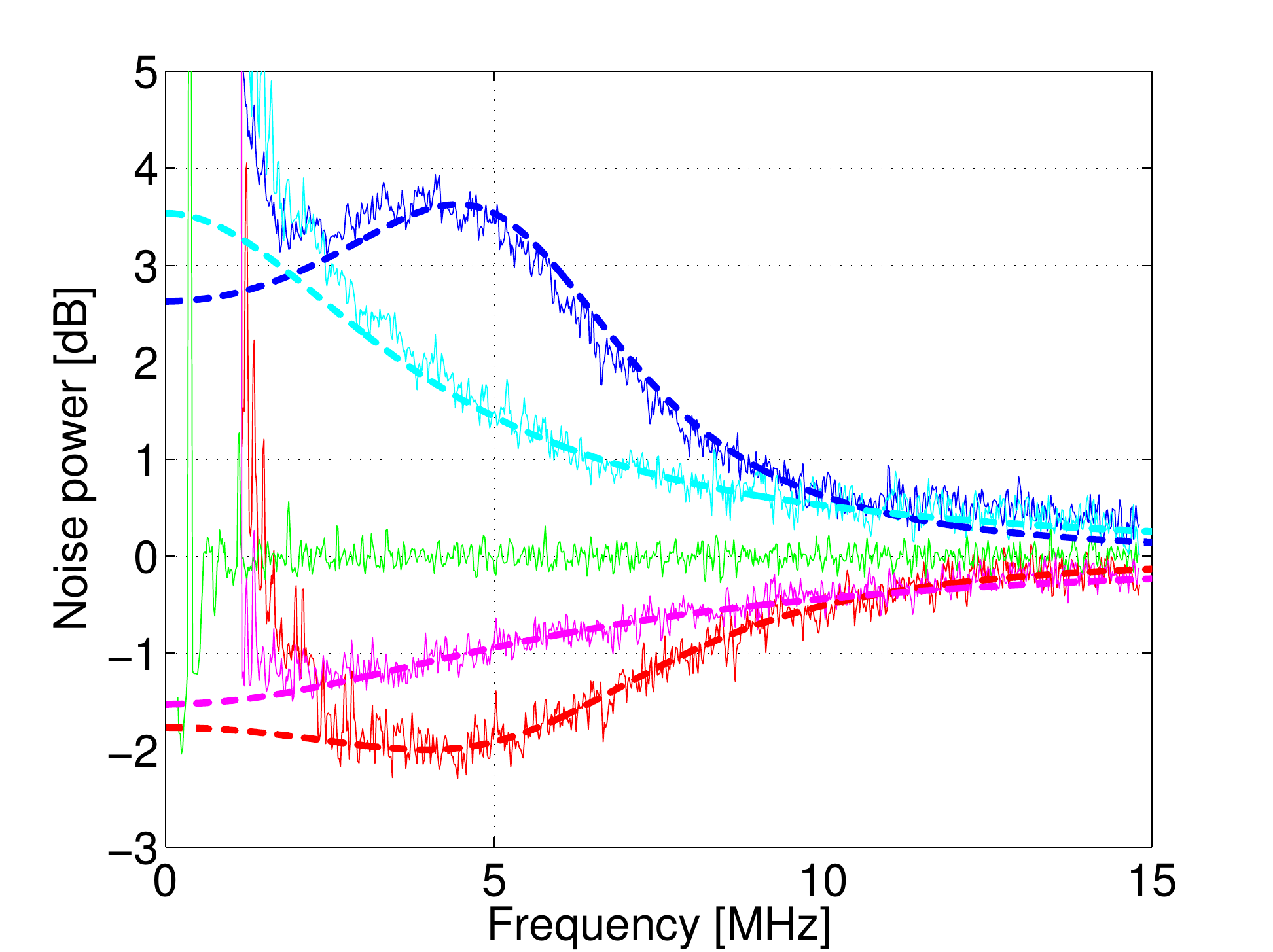}}
    \caption{Squeezing spectra for the empty-cavity destructive interference feedback, for different pump powers. $x = \sqrt{P^P_p/P^P_\text{th}}$. Red (magenta): closed-loop (open-loop) squeezing, Green: vacuum, blue (cyan): closed-loop (open-loop) anti-squeezing. The theoretical calculations are shown in dashed lines in corresponding colors.}\label{fig:emptycavity-feedback}
\end{figure}

We then configure the system to form the coherent feedback loop. We lock the feedback phase to the destructive interference condition as described above, resulting in the minimum circulating power. For the first measurement, the controller OPO pump remains off (the controller is simply an empty cavity). Figure \ref{fig:emptycavity-feedback} shows the squeezing spectra as well as the open loop spectra without feedback for comparison. We see a shift in the frequency of the closed loop spectra with respect to the open loop spectra. Both the squeezing and the anti-squeezing spectra in the feedback case are peaked at 4 MHz away from DC. The RF frequencies of the squeezing extrema stay nearly the same with increasing plant pump power, which agrees with the theory. We cannot measure the spectra for pump parameters over $x = 0.33$ due to instability of one of the phase lock loops at higher pump powers. However, in theory even when $x = 1.4$, the closed loop system poles stay in the negative half of the complex plane, which is significant to note as the open loop system becomes unstable for $x \ge 1$; thus, destructive interference feedback stabilizes the unstable open system. Figure \ref{fig:emptycavity-feedback} also compares the measured and the theoretically calculated squeezing spectra. For theoretical calculations we use the parameters shown in Table \ref{tab:param-emptycavity}, and we observe an excellent agreement between the measured and the spectra, which implies that the SLH quantum stochastic model successfully describes the all-optical coherent feedback OPO network.

An important effect of the feedback loop is that it increases the amount of squeezing achieved for a given plant pump power. For $x = 0.33$, we measure the maximum squeezing of the open loop configuration at slightly over 1 dB, while the empty-cavity coherent feedback configuration achieves nearly double the squeezing amount at the same of pump power.

We then turn on the pump power of the controller OPO, forming a true OPO network in coherent feedback configuration. We again set the coherent feedback phase to destructive interference. We define the two pump parameters as $x = 2 \epsilon / \gamma_T = \sqrt{P^P_p/P^P_\text{th}}$ for the plant OPO and $|y| = 2 \eta / \kappa_T = \sqrt{P^C_p/P^C_\text{th}}$ for the controller OPO, where $P_p^j, P_\text{th}^j$ are the pump powers and threshold pump powers for plant ($j = P$) and the controller ($j = C$), respectively. $y > 0$ corresponds to the case in which the controller and plant OPOs are both squeezing (or anti-squeezing) together, while $y < 0$ indicates reversed squeezing parity between the two OPOs (one squeezes while the other anti-squeezes). For the case where $x = 0.32, y = 0.10$, we observe that the peak squeezing and anti-squeezing values occur at 5 MHz. We compare with the empty-cavity destructive interference feedback case shown in Figure \ref{fig:emptycavity-feedback} (b). $x = 0.33$ with empty cavity feedback exhibits the peak squeezing near 4 MHz, and thus we see that having an active controller OPO further shifts the peak squeezing RF frequency by 1 MHz, as is predicted by the theory. Separately, we also note a slight increase in the anti-squeezing and squeezing peak values for $x = 0.32, ~y = 0.10$ when compared with $x = 0.32, y = 0$.

\begin{figure}[!tb]
	\centering
	\subfloat[$x = 0.32, y = 0.10$]{\includegraphics[width = 0.5\textwidth]{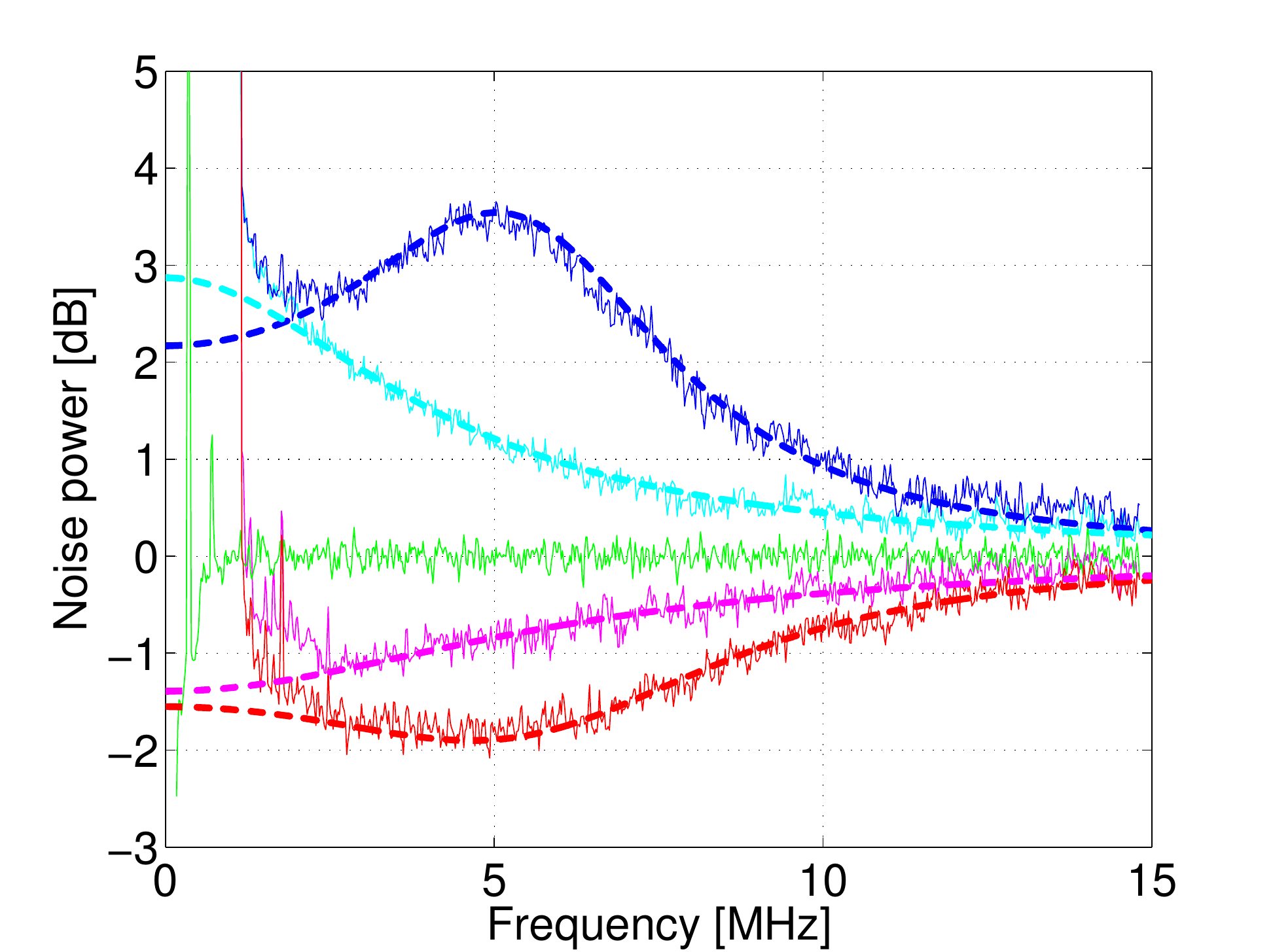}}
	\subfloat[$x = 0.32, y = -0.09$]{\includegraphics[width = 0.5\textwidth]{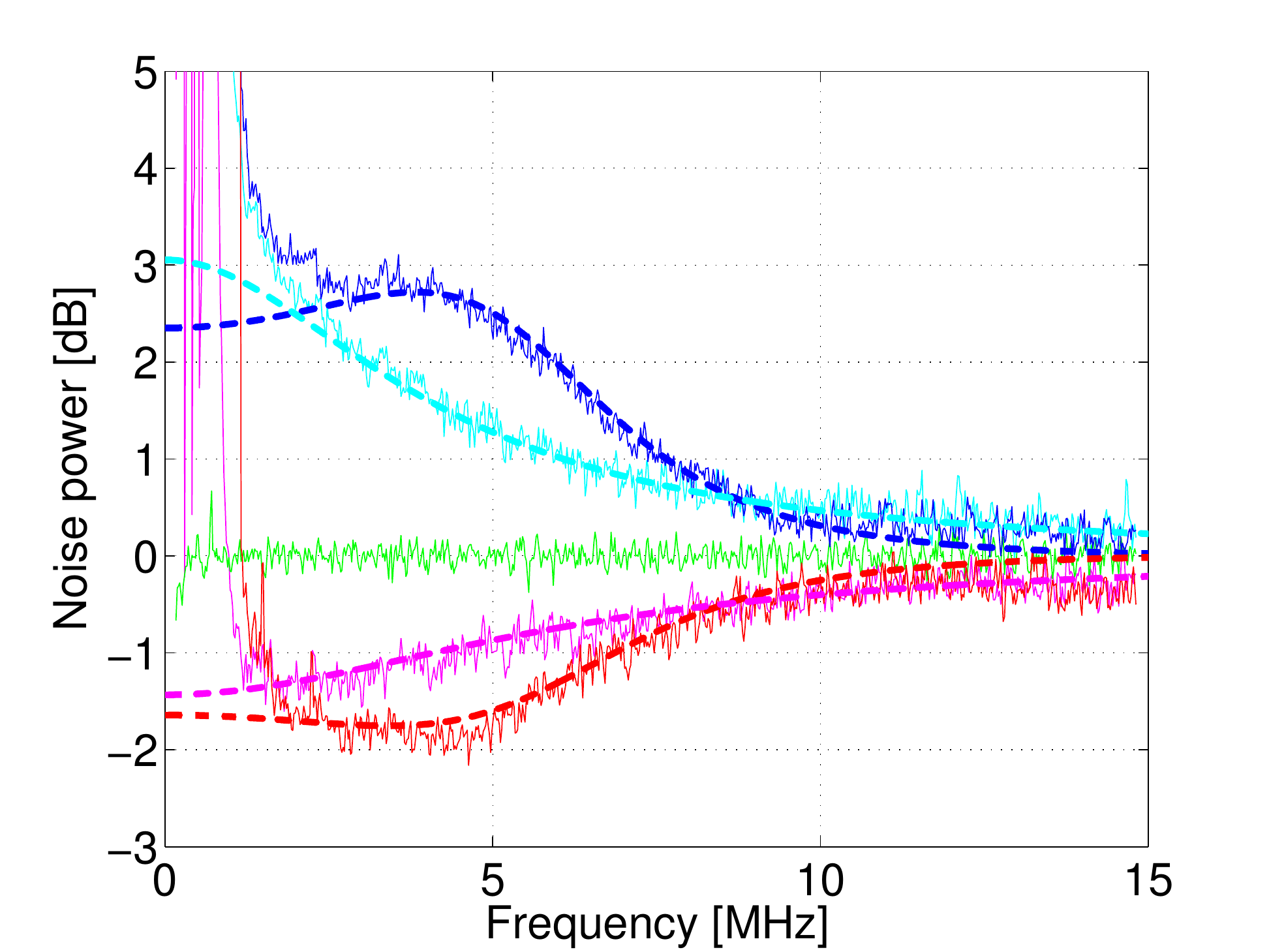}}
	\caption{Measured squeezing spectra (red) and anti-squeezing spectra (blue) of an OPO network for different pump parities. Here, $x = \sqrt{P^P_p/P^P_\text{th}}$ (plant pump) and $|y| = \sqrt{P^C_p/P^C_\text{th}}$ (controller pump). The open loop squeezing (magenta) and the anti-squeezing spectra (cyan) are shown for comparison. Also shown are the corresponding theoretical curves in dashed lines. }\label{fig:OPOnetwork-dest}
\end{figure}

Next, we reverse the pump parity between the plant and controller OPO. Figure \ref{fig:OPOnetwork-dest} (b) shows the measured squeezing spectra. This case displays reduced amounts of squeezing and anti-squeezing relative to the empty cavity controller case. Thus we observe that matching controller and plant pump parities enhances squeezing and anti-squeezing, whereas the case of opposite parities decreases squeezing and anti-squeezing.

\begin{figure}[!tb]
	\centering
	\subfloat[16 MHz detuned controller OPO cavity]{\includegraphics[width=0.5\textwidth]{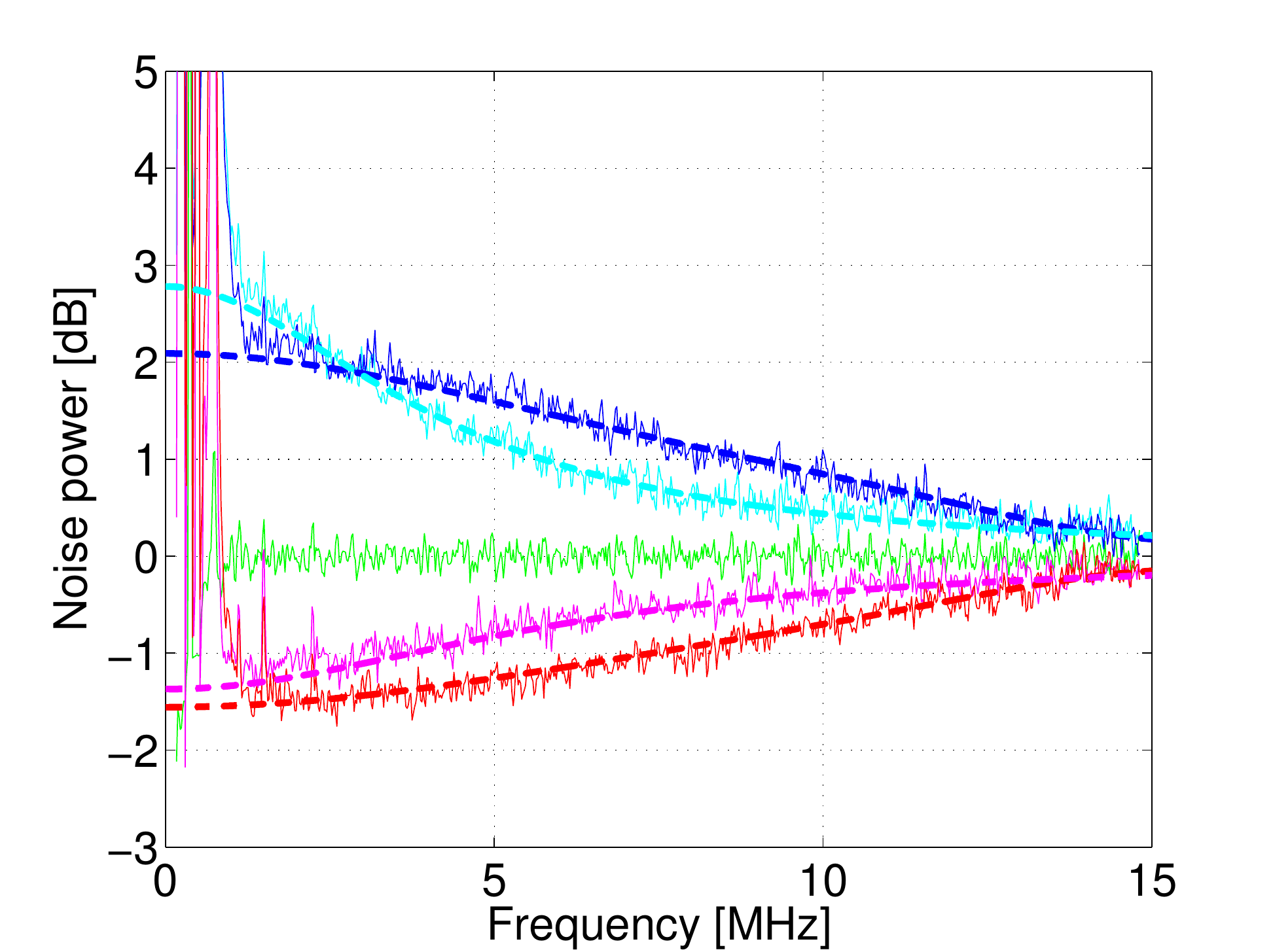}}
	\subfloat[Circulating power vs. feedback phase]{\includegraphics[width=0.5\textwidth]{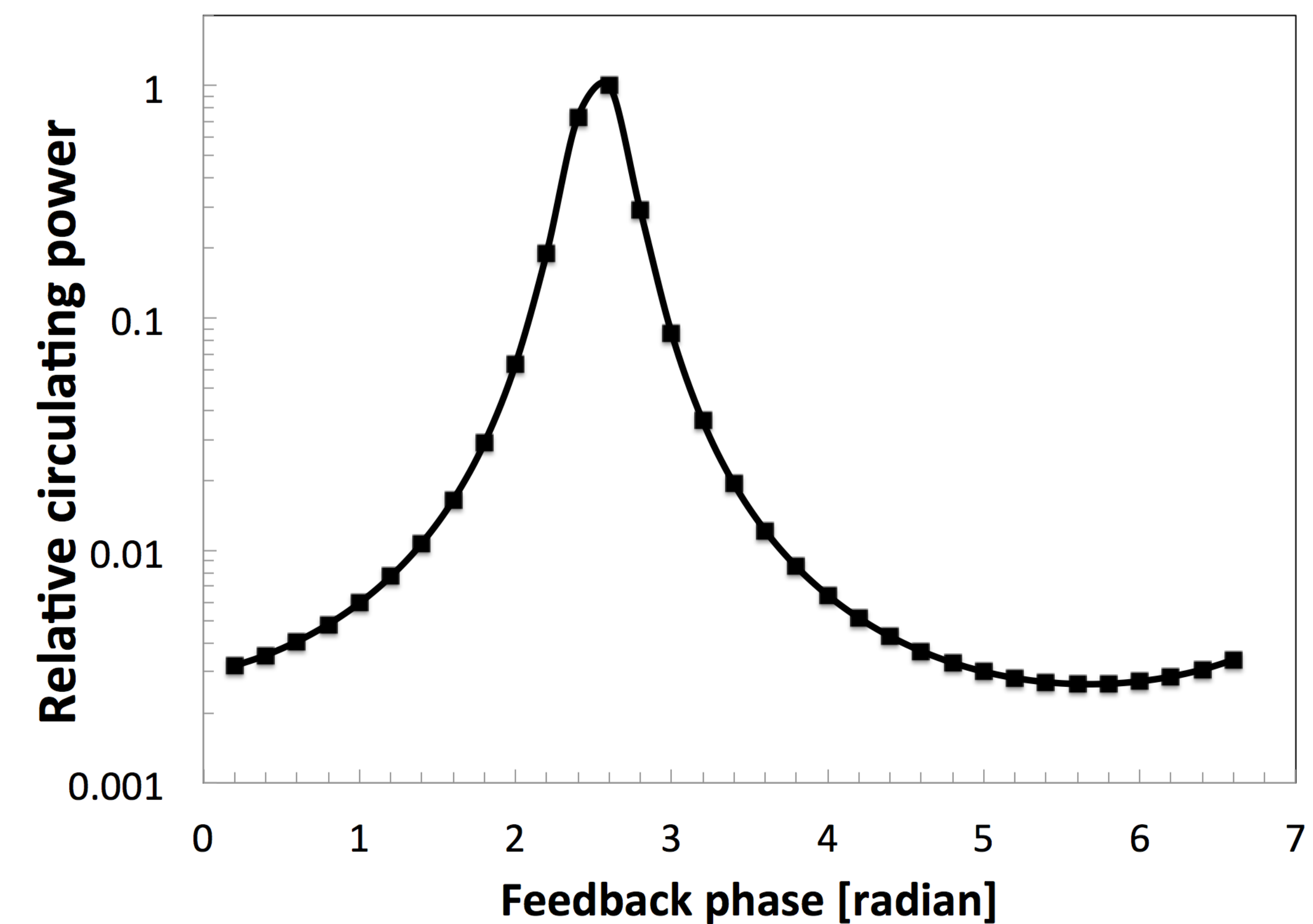}}
	\caption{Squeezing (red) and anti-squeezing (blue) spectra for a 16 MHz detuned controller OPO cavity (left). Here, $x = 0.29$ (plant pump) and $y = 0$ (controller pump). The open loop squeezing (magenta) and the anti-squeezing (cyan) spectra are shown for comparison. Theoretical curves for corresponding squeezing spectra are shown in dotted lines. The theoretical relative circulating power as a function of feedback phase (right). }\label{fig:detuning}
\end{figure}

We also measure the spectra for the constructive interference feedback configuration, and we briefly describe the results as the following. The constructive interference coherent feedback configuration yields significantly reduced squeezing spectra, accompanied by the addition of significant amounts of low frequency noise as the two OPO network becomes less stable and noises in the phase lock loops increase. This feedback configuration substantially reduces the lasing threshold pump power of the OPO network (note that our locking schemes fail near and above this threshold); in other words, a small amount of pump power ($\sim$10's of mW) injected into the plant OPO is now enough to push the combined system above threshold. From a dynamical systems perspective, this configuration illustrates that the stability of the controlled system (the plant, stable by itself at the same input pump power) can be modified (in this case, made unstable) with the addition of coherent feedback.

Finally, we investigate the effects of detuning the controller cavity resonance from the optical drive frequency. The cavity detuning is accomplished by changing the driving RF frequency for the AOM used to shift the locking beam frequency (nominally by one free-spectral range). The coherent feedback phase is set to the destructive interference condition. We detune the cavity frequency by 16 MHz and observe that the squeezing and the anti-squeezing spectra broaden, as predicted by theory. Figure \ref{fig:detuning} shows the result. The broadening of the squeezing and the anti-squeezing spectra increases as the controller detuning is increased, yielding enhanced squeezing at larger frequency detunings from the optical drive frequency. We note that when the controller cavity is detuned from the optical drive frequency, the coherent feedback phase $\phi$ that corresponds to the destructive interference condition is no longer 0 or $\pi$. In order to compare the experimental result with the theory, it is important to find out the correct feedback phase $\phi$ for the destructive feedback. For this, we calculated the steady-state circulating power as a function of the feedback phase from our theory (the result is plotted in Figure \ref{fig:detuning} (b)). Figure \ref{fig:detuning} (a) shows the theoretical squeezing and anti-squeezing curves using $\phi = 5.6$ radians, which corresponds to the calculated minimum circulating power. The theory agrees with the experimental result, which is a proof that our theory on photon numbers explains the experimental result well. We also note that the effect of the broadening of squeezing spectra diminishes for a large controller cavity detunings away from the driving frequency, as transmission into the controller cavity (and hence its interaction with the plant in the feedback loop) is reduced. When we detune the controller cavity far off the driving frequency, the squeezing and the anti-squeezing spectra converge to the trivial feedback case without the controller cavity.

\section{Conclusions and discussion}

In this work, we have investigated the squeezing spectra of a two OPO coherent feedback network, and found that controller OPO parameters such as the pump phase, pump power and cavity detuning significantly alter the properties of the squeezing spectra produced at the plant OPO output. We also found that the coherent feedback phase had a large impact in determining both squeezing performance and overall system stability. To validate our experimental results, we developed the relevant theory to explain the detailed behavior of the OPO network system based on the SLH framework. The agreement between our theory and measured squeezing spectra was excellent, implying that this theoretical framework is well-suited for use in designing of the functionality OPO squeezing networks and similar coherent quantum feedback systems. Specifically, by exploiting the linear dynamics of our system in the sub-threshold regime we were able to apply well-established methods of linear dynamical systems theory (the ABCD-framework suited for linear QSDEs) to analyze our system.

We presented experimental squeezing spectra in various feedback configurations. As mentioned above, the constructive interference feedback configuration reduced the lasing threshold of the combined two OPO system (from the perspective of the plant OPO, connecting its two input-output ports in a constructive manner essentially reduced the roundtrip propagation loss of the OPO), and made the system difficult to stabilize. In contrast, the destructive interference feedback configuration increased system stability, and enabled us to study the dynamic interaction between the plant and controller OPO cavity modes. The frequency dependence of the squeezing spectra can be understood from the frequency dependent transmission characteristics of the two OPO cavities: we found that the peak transmission frequency coincided with the peak squeezing frequency. This suggests that the feedback adds additional frequency dependent amplification or deamplification that is dependent on the frequency-specific phase of the feedback. Some prominent effects of this, as we showed experimentally, were that we could tailor the system properties to shift the frequency of maximum squeezing and anti-squeezing away from the optical drive frequency, as well as broaden spectrum over a wider band of frequencies, which may play a significant role in future quantum engineering applications. We note that this table-top optical setup required substantial classical (feedback) control to stabilize and lock our various components, but this enabled us to have a tunable and flexible system. For specific applications, a less tunable, but more stable system could be implemented, for example in a nanophotonic setting \cite{fuerst2010lino3opo, painter2013optmechsqueez}. Finally, we would like to mention that the theory predicts additional interesting features that should be present in this system, such as limit cycle dynamical regimes that occur when the two OPO network is driven above threshold; the study of such behavior in the above threshold regime will be a topic for future work.

\section*{Acknowledgments}
This work has been supported by AFOSR (FA9550-11-1-0238) and DARPA (N66001-11-1-4106). 
Orion Crisafulli would like to acknowledge support from the National Defense Science and Engineering Graduate Fellowship and National Science Foundation Graduate Fellowship programs while a graduate student at Stanford University, where this work was done, before joining MIT Lincoln Laboratory, where he is presently employed.
Nikolas Tezak is supported by a Stanford Graduate Fellowship and a Simons Foundation Math+X Fellowship.

Sandia National Laboratories is a multi-program laboratory managed and operated by Sandia Corporation, a wholly owned subsidiary of Lockheed Martin Corporation, for the U.S. Department of Energy’s National Nuclear Security Administration under contract DE-AC04-94AL85000.


\begin{thebibliography}{squeezing}
%
\bibitem{wiseman1994all} H. M. Wiseman and G. J. Milburn, ``All-optical versus electro-optical quantum-limited feedback," \pra {\bf 49} (5), 4110 (1994).

\bibitem{lloyd2000coherent} S. Lloyd, ``Coherent quantum feedback," \pra {\bf 62} (2), 022108 (2000).

\bibitem{yanagisawa2003transfer} M. Yanagisawa, and H. Kimura, ``Transfer function approach to quantum control-part I: Dynamics of quantum feedback systems," IEEE Trans.\ Automatic Control {\bf 48} (12), 2107--2120 (2003).

\bibitem{gough2009series} J. Gough and M. R. James, ``The series product and its application to quantum feedforward and feedback networks," IEEE Trans. on Automatic Control {\bf 54} (11), 2530--2544 (2009).

\bibitem{belavkin1992quantum} V. P. Belavkin, ``Quantum stochastic calculus and quantum nonlinear filtering," J. Multivariate Analysis {\bf 42} (2), 171--201 (1992).

\bibitem{bouten2007introduction} L. Bouten, R. Van Handel, and M. R. James, ``An introduction to quantum filtering," SIAM J. Control and Optimization {\bf 46} (6), 2199--2241 (2007).

\bibitem{mabuchi2011lowpower} H. Mabuchi, ``Coherent-feedback control strategy to suppress spontaneous switching in ultralow power optical bistability," \apl {\bf 98} (19), 193109 (2011).

\bibitem{kerckoff2010QEC} J. Kerckhoff et. al., ``Designing Quantum Memories with Embedded Control: Photonic Circuits for Autonomous Quantum Error Correction," \prl {\bf 105} (19), 040502 (2010).

\bibitem{hudson1984quantum} R. L. Hudson and K. R. Parthasarathy, ``Quantum Ito's formula and stochastic evolutions," Comm.\ in Math.\ Phys.\, {\bf 93} (3), 301--323 (1984).

\bibitem{gardiner1985input} C. W. Gardiner and M. J. Collett, ``Input and output in damped quantum systems: Quantum stochastic differential equations and the master equation," \pra {\bf 31} (6), 3761 (1985).

\bibitem{gardiner1985cascaded} C. W. Gardiner, ``Driving a quantum system with the output field from another driven quantum system," \prl {\bf 70} (15), 2269--2272 (1993).

\bibitem{carmichael1993cascaded} H. J. Carmichael, ``Quantum Trajectory Theory for Cascaded Open Systems,'' \prl {\bf 70} (15), 2273--2276 (1993).

\bibitem{mabuchi2008coherent} H. Mabuchi, ``Coherent-feedback quantum control with a dynamic compensator," \pra {\bf 78} (3), 032323 (2008).

\bibitem{iida2012experimental} S. Iida, M. Yukawa, H. Yonezawa, N. Yamamoto, and A. Furusawa, ``Experimental demonstration of coherent feedback control on optical field squeezing," IEEE Trans. Automatic Control {\bf 57} (8), 2045--2050 (2012).

\bibitem{kerckhoff2012multivibrator} J. Kerckhoff, and K. W. Lehnert, ``Superconducting Microwave Multivibrator Produced by Coherent Feedback,'' \prl {\bf 109} (15), 153602 (2012).

\bibitem{ralph2003qkd} T. C. Ralph, ``Quantum Key Distribution with Continuous Variables in Optics" in {\it Quantum Information with Continuous Variables,} Springer, (2003).

\bibitem{vahlbruch2007quantum} H. Vahlbruch, S. Chelkowski, K. Danzmann, and R. Schnabel. ``Quantum engineering of squeezed states for quantum communication and metrology." New Journal of Physics {\bf 10} (9), 371 (2007).

\bibitem{collett1984squeezing} M. J. Collett and C. W. Gardiner, ``Squeezing of intracavity and traveling-wave light fields produced in parametric amplification," \pra {\bf 30} (3), 1386 (1984).

\bibitem{slusher1985observation} R. E. Slusher, L. W. Hollberg, B. Yurke, J. C. Mertz, and J. F. Valley, ``Observation of squeezed states generated by four-wave mixing in an optical cavity," \prl {\bf 55} (22), 2409-2412 (1985).

\bibitem{wu1987squeezed} L.-A. Wu, M. Xiao, and H. J. Kimble. ``Squeezed states of light from an optical parametric oscillator,"  \josab {\bf 4} (10), 1465-1475 (1987).

\bibitem{caves1981quantum} C. M. Caves, ``Quantum-mechanical noise in an interferometer," \prd {\bf 23} (8), 1693 (1981).

\bibitem{thorne1987300} K. S. Thorne, S. Hawking, and W. Israel, {\it 300 years of gravitation,} Cambridge University Press, Cambridge, England (1987).

\bibitem{kimble2001conversion} H. J. Kimble, Y. Levin, A. B. Matsko, and K. S. Thorne, and S. P. Vyatchanin, ``Conversion of conventional gravitational-wave interferometers into quantum nondemolition interferometers by modifying their input and/or output optics,"" \prd {\bf 65} (2), 022002 (2001).

\bibitem{gea1987squeezed} J. Gea-Banacloche and G. Leuchs. ``Squeezed states for interferometric gravitational-wave detectors," \jmo {\bf 34} (6-7), 793--811 (1987).

\bibitem{vahlbruch2006coherent} H. Vahlbruch, S. Chelkowski, B. Hage, A. Franzen, K. Danzmann, and R. Schnabel, ``Coherent control of vacuum squeezing in the gravitational-wave detection band,'' \prl {\bf 97} (1) 011101 (2006).

\bibitem{cerf2005qkd} N.J. Cerf, J. Clavareau, C. Macchiavello, and J. Roland, ``Quantum entanglement enhances the capacity of bosonic channels with memory,'' \pra {\bf 72} 042330 (2005).

\bibitem{zan2012opacascade} Z. Yan, X. Jia, X. Su, Z. Duan, C. Xie, and K. Peng, ``Cascaded entanglement enhancement,'' \pra {\bf 85} (4), 040305 (2012).

\bibitem{eberle2010quantum} T. Eberle, S. Steinlechner, J. Bauchrowitz, V. H{\"a}ndchen, H. Vahlbruch, M. Mehmet, H. M{\"u}ller-Ebhardt, and R. Schnabel, Roman, ``Quantum enhancement of the zero-area Sagnac interferometer topology for gravitational wave detection,'' \prl {\bf 25} 251102 (2010).

\bibitem{gough2009enhancement} J. E. Gough and S. Wildfeuer, ``Enhancement of field squeezing using coherent feedback," \pra {\bf 80} (4), 042107 (2009).

\bibitem{zhang2012qfnreview} G. Zhang, and M. R. James, ``Quantum Feedback Networks and Control: A Brief Survey," Chinese Science Bulletin {\bf 57} (18), 2200 -- 2214 (2012).

\bibitem{drever1983laser} R. W. P. Drever, J. L. Hall, and F. V. Kowalski, J. Hough, G. M. Ford, A. J. Munley, and H. Ward, ``Laser phase and frequency stabilization using an optical resonator," \apb {\bf 31} (2), 97--105 (1983).

\bibitem{yuen1980optical} H. P. Yuen and J. Shapiro, ``Optical communication with two-photon coherent states--Part III: quantum measurements realizable with photoemissive detectors," IEEE TRans. on Information Theory {\bf 1} (26), 78--92 (1980).

\bibitem{yuen1983noise} H. P. Yuen and V. W. S. Chan, ``Noise in homodyne and heterodyne detection," \ol {\bf 8} (3), 177--179 (1983).

\bibitem{breitenbach1997measurement} G. Breitenbach, S. Schiller, and J. Mlynek, ``Measurement of the quantum states of squeezed light," Nature {\bf 387}, 471--475 (1997).

\bibitem{takeno2007observation} Y. Takeno, M. Yukawa, H. Yonezawa, and A. Furusawa, ``Observation of-9 dB quadrature squeezing with improvement of phase stability in homodyne measurement," \opex {\bf 15} (7), 4321--4327 (2007).


\bibitem{fuerst2010lino3opo} J.U. Fürst, D. V. Strekalov,D. Elser,A. Aiello,U. L. Andersen, C. Marquardt, and G. Leuchs, ``Low-Threshold Optical Parametric Oscillations in a Whispering Gallery Mode Resonator,'' \prl {\bf 105} (26), 263904 (2010).

\bibitem{painter2013optmechsqueez} A. H. Safavi-Naeini, S. Groeblacher, J. T. Hill, J. Chan, M. Aspelmeyer, and O. Painter, ``Squeezing of light via reflection from a silicon micromechanical resonator,''  arXiv:1302.6179 [quant-ph]


%
%
%
%
%
%
%
%
%

\end{thebibliography}
\end{document}